# Scalability Evaluation of Iterative Algorithms Used for Supercomputer Simulation of Physical processes

Nadezhda A. Ezhova, Leonid B. Sokolinsky
South Ural State University
Chelyabinsk, Russian Federation
EzhovaNA@susu.ru, leonid.sokolinsky@susu.ru

*Abstract* — The paper is devoted to the development of a methodology for evaluating the scalability of compute-intensive iterative algorithms used in simulating complex physical processes on supercomputer systems. The proposed methodology is based on the BSF (Bulk Synchronous Farm) parallel computation model, which makes it possible to predict the upper scalability bound of an iterative algorithm in early phases of its design. The BSF model assumes the representation of the algorithm in the form of operations on lists using high-order functions. Two classes of representations are considered: BSF-M (Map BSF) and BSF-MR (Map-Reduce BSF). The proposed methodology is described by the example of the solution of the system of linear equations by the Jacobi method. For the Jacobi method, two iterative algorithms are constructed: Jacobi-M based on the BSF-M representation and Jacobi-MR based on the BSF-MR representation. Analytical estimations of the speedup, parallel efficiency and upper scalability bound are constructed for these algorithms using the BSF cost metrics on multiprocessor computing systems with distributed memory. An information about the implementation of these algorithms in C++ language using the BSF program skeleton and MPI parallel programming library are given. The results of large-scale computational experiments performed on a cluster computing system are demonstrated. Based on the experimental results, an analysis of the adequacy of estimations obtained analytically by using the cost metrics of the BSF model is made.

## I. Introduction

The concept of "Industry 4.0" considers smart factories as data-driven and knowledge enabled enterprise intelligence. In such kind of factory, manufacturing processes and final products are accompanied by virtual models – Digital Twins [1]. *The Digital Twin* (DT) supports virtual models of real equipment, industrial process, and final products. The Digital Twin provides methods of analysis of data from diverse types of sensors installed on the objects for tuning and actualization of their virtual state. As a rule, the physical model is described by a system of differential equations for which it is impossible to find an analytical solution. In this case, the numerical methods for solving systems of differential equations reducing the solution of the original problem to the solution of a system of linear algebraic equations are used. Linear equation systems used to model complex technical devices and processes usually have high computational complexity. To solve such large-scale systems at an acceptable time, one must use scalable parallel algorithms intended for multiprocessor computing systems with distributed memory. When one creates parallel algorithms for large multiprocessor systems, it is important at an early stage of the algorithm design (before coding) to obtain an analytical estimation of its scalability. For this purpose, one can use various models of parallel computation [2]. Nowadays, a large number of different parallel computation models are known. The most famous models among them are PRAM [3], BSP [4] and LogP [5]. Each of these models generated a large family of parallel computation models, which extend and generalize the parent model (see, e.g., [6]–[8]). The problem of developing new parallel computation models is still important today. The reason is that it is impossible to create a parallel computation model, which is good in all respects. To create a good parallel computation model, the designer must restrict the set of target multiprocessor architectures and class of algorithms. In paper [9], the parallel computation model BSF (Bulk Synchronous Farm) intended for cluster computing systems and iterative algorithms was proposed. The BSF model is an extension of the BSP model which is based on the SPMD (Single-Program-Multiple-Data) programming model [10], [11] and the master-worker (master-slave) framework [12]. The BSF model makes it possible to predict the upper scalability bound of an iterative algorithm with great accuracy before coding. Some examples of using the BSF model are given in [13], [14]. The purpose of this article is to study the scalability of iterative algorithms used in supercomputer simulation of physical processes on multiprocessor systems with distributed memory by using the BSF parallel computation model. As candidates, we chosen two different implementations of the Jacobi method used for the solution of large-scale linear equation systems. The rest of the article is organized as follows. Section 1 provides a brief overview of the BSF parallel computation model. Section 2 is devoted to describing the representation of iterative numerical algorithms using high-order functions *Map* and *Reduce* defined in the Bird–Meertens formalism. For these representations, a general approach to parallelization of computations is proposed. In Section 3, the cost metrics of the BSF model are presented. These cost metrics allow us to obtain analytical estimations for speedup, parallel efficiency, and upper bound of scalability of a parallel algorithm. Section 4 gives a formal description of the Jacobi method for solving linear equation systems. Section 5 describes the Jacobi-M algorithm implementing the Jacobi method in the form of operations on lists using the high-order function *Map*. In Section 6, we derive the analytical estimations of speedup, parallel efficiency and upper scalability bound of the Jacobi-M algorithm using the BSF model cost metrics. Section 7 describes the Jacobi-MR algorithm implementing the Jacobi method in the form of operations on lists using the high-order functions *Map* and *Reduce*. In Section 8, we obtain the analytical estimations of speedup, parallel efficiency and upper scalability bound of the Jacobi-MR algorithm using the BSF model cost metrics. Section 9 provides a description of the implementations of the Jacobi-M and Jacobi-MR algorithms in C++ language using the BSF algorithmic skeleton and the MPI parallel programming library. A comparison of the results obtained analytically and experimentally is given. In conclusion, the obtained results are summarized and directions for further research are outlined.

## II. BSF PARALLEL COMPUTATION MODEL

The idea of the *parallel computation model BSF* (Bulk Synchronous Farm) was originally proposed in [9]. In this section, we present a brief description of an extension of the original model. A *BSF-computer* consists of a collection of homogeneous computing nodes with private memory connected by a communication network delivering messages among the nodes. There is just one node called the master-node in a BSF-computer. The rest of the nodes are the worker-nodes. A BSF-computer must include at least one master-node and one worker-node. Let us denote the number of worker-nodes by $K$. We assume that $K \geq 1$. A BSF-computer utilizes the SPMD programming model. A BSF-program consists of sequences of macrosteps and global barrier synchronizations performed by the master and all the workers. Each macro-step is divided into two sections: the master section and the worker section. The master section includes instructions performed by only the master. A worker section includes instructions performed by only the workers. The sequential order of the master section and the worker section within the macro-step is not important. All the worker nodes operate on the same data array, but the base address of the data assigned to the worker-node for processing is determined by the logical number of this node. A BSF-program includes the following sequential sections: initialization; iterative process; finalization. *Initialization* is a macro-step in which the master and workers read or generate input data. Initialization is followed by barrier synchronization. The *iterative process* repeatedly performs its body until the stopping criterion checked by the master becomes true. In the *finalization* macro-step, the master outputs the results and ends the program.

The body of the iterative process includes the following macro-steps:
1) sending orders (from master to workers);
2) processing orders (by workers);
3) receiving results (from workers to master);
4) evaluating the results (by master).

In the first macro-step, the master sends the same orders to all workers. Then, the workers execute the received orders (the master is idle at that time). All the workers execute the same program code but operate on different data with a base address which depends on the worker-node number. Therefore, all workers spend the same amount of time on calculation. There are no data transfers between nodes during order processing. In the third macro-step, all workers send the results to the master. Next, global barrier synchronization is performed. During the fourth macro-step, the master evaluates the results it has received. The workers are idle at this time. After evaluation of the results, the master checks the stopping criterion. If the stopping criterion is true, then the iterative process is finished, otherwise the iterative process is continued.

The BSF model includes the following main cost parameters within one iteration:

$K$ – the number of worker-nodes;
$t_s$ – the time that the master-node is engaged in sending one order to one worker-node, excluding latency;
$t_w$ – the time a BSF-computer with one worker-node needs to perform one order;
$t_R$ – the total time that the master-node is engaged in receiving the results from all worker-nodes, excluding latency;
$t_p$ – the total time that the master-node is engaged in evaluating the results received from all worker-nodes and checking the stopping criterion;
$L$ – the latency (the time it takes to send a 1-byte message).

Based on these parameters, the BSF model allows us to obtain analytical estimations of speedup, parallel efficiency and an upper scalability bound of the algorithm. However, the final form of the equations giving such estimations depends on the chosen class of the representation of the algorithm in the form of operations on lists using high-order functions. The BSF model provides two classes of representations: BSF-M and BSF-MR. In the *BSF-M representation*, the implementation of the algorithm is performed using the high-order function *Map*. In the *BSF-MR representation*, the implementation of the algorithm is performed using the high-order function *Map*. In the BSF-MR representation, in addition to *Map*, the high-order function *Reduce* is used. Let us briefly consider high-order functions *Map* and *Reduce* in the following section.

### III. OPERATIONS ON LISTS

To obtain analytical estimations of an algorithm using the BSF model metrics, this algorithm must be represented in the form of operations on lists using the high-order functions *Map* and *Reduce* defined in the Bird–Meertens formalism [15].

The higher-order function *Map* applies the given function $F: \mathbb{A} \to \mathbb{B}$ to each element of the given list $[a_1, \ldots, a_l]$ and returns a list of results in the same order:

$$Map(F, [a_1, \ldots, a_l]) = [F(a_1), \ldots, F(a_l)]. \quad (1)$$

The higher-order function *Reduce* reduces the given list $[b_1, \ldots, b_l]$ to a single value by iteratively applying the given binary associative operation $\oplus: \mathbb{B} \times \mathbb{B} \to \mathbb{B}$ to each pair of elements:

$$Reduce(\oplus, [b_1, \ldots, b_l]) = b_1 \oplus \ldots \oplus b_l. \quad (2)$$

Let us denote the operation of lists concatenation by the symbol $++$. From equation (1) follows:

$$\begin{aligned} Map(F, [a_1, \ldots, a_l]) = Map(F, [a_1, \ldots, a_m]) ++ \\ ++ \ldots ++ Map(F, [a_{1+(k-1)m}, \ldots, a_l]) \end{aligned} \quad (3)$$

for any $m, k, l \in \mathbb{N}$, such that $l = k \cdot m$. This means that the function *Map* can be executed in parallel on $k$ sublists of length $m$ making up the original list $[a_1, \ldots, a_l]$ after that the resulting lists must be concatenated (see algorithm 1). The iterations of **pardo** loop can be executed in parallel because they are mutually data-independent. From equations (1) and (2) it follows, that:

$$\begin{aligned} Reduce(\oplus, Map(F, [a_1, \ldots, a_l])) = \\ = Reduce\left(\oplus, \begin{bmatrix} Reduce(\oplus, Map(F, [a_1, \ldots, a_m])), \ldots, \\ Reduce(\oplus, Map(F, [a_{(k-1)m+1}, \ldots, a_l])) \end{bmatrix}\right) \end{aligned} \quad (4)$$

---

**Algorithm 1.** Parallel execution of *Map* function
**Input**: List $[a_1, \ldots, a_l]$ of length $l = k \cdot m$.
**Output**: List $[b_1, \ldots, b_l]$, there $b_i = F(a_i)$.
**begin**
  **for** $j = 1$ **to** $k$ **pardo**
    $[b_{1+m(j-1)}, \ldots, b_{j \cdot m}] = Map(F, [a_{1+m(j-1)}, \ldots, a_{j \cdot m}])$
  **end for**
  **return** $[b_1, \ldots, b_l]$
**end**

---

for any $m, k, l \in \mathbb{N}$, such that $l = k \cdot m$. This means that the *Map/Reduce* function composition can be executed in parallel on $k$ sublists of length $m$ making up the original list $[a_1, \ldots, a_l]$ after that the resulting lists must be concatenated (see algorithm 2).

---

**Algorithm 2.** Parallel execution of *Map/Reduce* function *composition*
**Input**: List $[a_1, \ldots, a_l]$ of length $l = k \cdot m$.
**Output**:
Value $B = Reduce(\oplus, Map(F, [a_1, \ldots, a_l]))$.
**begin**
  **for** $j = 1$ **to** $k$ **pardo**
    $[b_{1+m(j-1)}, \ldots, b_{j \cdot m}] = Map(F, [a_{1+m(j-1)}, \ldots, a_{j \cdot m}])$
    $B_j = e$
    **for** $i = 1 + m(j-1)$ **to** $j \cdot m$
      $B_j = B_j \oplus b_i$
    **end for**
  **end for**
  $B = e$
  **for** $j = 1$ **to** $k$ **pardo**
    $B = B \oplus B_j$
  **end for**
  **return** $B$
**end**

---

Here, $e \in \mathbb{B}$ denotes an identity element for the operation $\oplus$:

$$b \oplus e = b$$

for any element $b \in \mathbb{B}$.

### IV. BSF MODEL COST METRICS

BSF model provides the following cost metrics for analytical estimation of speedup, parallel efficiency and upper bound of scalability of a parallel algorithm. For the BSF-M representation, the speedup as a function of $K$ is estimated by the equation:

$$a_{BSF-M}(K) = \frac{K(2L + t_s + t_R + t_p + t_w)}{K^2(2L + t_s) + K(t_R + t_p) + t_w}. \quad (5)$$

The parallel efficiency of an algorithm as a function of $K$ can be estimated as follows:

$$e_{BSF-M}(K) = \frac{2L + t_s + t_R + t_p + t_w}{K^2(2L + t_s) + K(t_R + t_p) + t_w}. \quad (6)$$

The upper bound of scalability of an algorithm is estimated by the following equation:

$$P_{BSF-M} = \sqrt{\frac{t_w}{2L + t_s}}. \quad (7)$$

Here, $P_{BSF-M}$ denotes the number of worker-nodes, for which the maximum speedup of the BSF-M representation of the algorithm is achieved.

For BSF-MR representation, three additional parameters are introduced:

$t_r$ – the time it takes to send the result obtained by one worker-node to master-node;
$t_a$ – the time it takes to perform one operation $\oplus$, which is a parameter of the high-order function *Reduce*;
$l$ – the length of the *Reduce* list.

In this case, speedup is calculated by the equation:

$$a_{BSF-MR}(K) =$$
$$= \frac{2L + t_s + t_r + t_p + t_w + lt_a}{K(2L + t_s + t_r + t_a) + (t_w + lt_a)/K - t_a + t_p}. \quad (8)$$

The parallel efficiency can be estimated as follows:

$$e_{BSF-MR}(K) =$$
$$= \frac{2L + t_s + t_r + t_p + t_w + lt_a}{K^2(2L + t_s + t_r + t_a) + K(t_p - t_a) + t_w + lt_a}. \quad (9)$$

The upper bound of scalability is estimated by the following equation:

$$P_{BSF-MR} = \sqrt{\frac{t_w + lt_a}{2L + t_s + t_r + t_a}}. \quad (10)$$

## V. Jacobi method

The Jacobi iterative method [16] is an algorithm for determining the solutions of a diagonally dominant system of linear equations. This method was first described by the German mathematician Carl Gustav Jacob Jacobi in [17].

Practical application of the Jacobi method became possible with the advent of computers [18]. Nowadays, the interest of researchers in the Jacobi method is still preserved (see, e.g., [19], [20]). Let us give a brief description of the Jacobi method.

Let a joint system of linear equations in a matrix form be given in Euclidean space $\mathbb{R}^n$:

$$Ax = b, \quad (11)$$

where

$$A = \begin{pmatrix} a_{11} & \cdots & a_{1n} \\ \vdots & \ddots & \vdots \\ a_{n1} & \cdots & a_{nn} \end{pmatrix};$$
$$x = (x_1, \ldots, x_n);$$
$$b = (b_1, \ldots, b_n).$$

It is assumed that $a_{ii} \neq 0$ for all $i = 1, \ldots, n$. Let us define the matrix

$$C = \begin{pmatrix} c_{11} & \cdots & c_{1n} \\ \vdots & \ddots & \vdots \\ c_{n1} & \cdots & c_{nn} \end{pmatrix}$$

in the following way:

$$c_{ij} = \begin{cases} -\dfrac{a_{ij}}{a_{ii}}, \forall j \neq i; \\ 0, \forall j = i. \end{cases}$$

Let us define the vector $d = (d_1, \ldots, d_n)$ as follows: $d_i = b_i / a_{ii}$.

The Jacobi method of finding an approximate solution of system (11) consists of the following steps:

Step 1. $k := 0$; $x^{(0)} := d$.

Step 2. $x^{(k+1)} := Cx^{(k)} + d$.

Step 3. If $\left\| x^{(k+1)} - x^{(k)} \right\|^2 < \varepsilon$, got to Step 5.

Step 4. $k := k + 1$; go to Step 2.

Step 5. Stop.

In the Jacobi method, an arbitrary vector $x^{(0)}$ can be taken as the initial approximation. In Step 1, the initial approximation $x^{(0)}$ is assigned by the vector $d$. In Step 3, the Euclidean norm $\|\cdot\|$ is used in the stopping criterion.

The diagonal dominance of the matrix $A$ is a sufficient condition for the convergence of the Jacobi method, i.e.:

$$|a_{ii}| \geq \left( \sum_{j=1}^{n} |a_{ij}| \right) - |a_{ii}|$$

for all $i = 1, \ldots, n$, and at least one inequality is strict. Diagonally dominant matrices often arise in applications. In this case, system (11) has a unique solution for any right-hand sides.

## VI. Jacobi-M algorithm

The Jacobi-M algorithm implements the Jacobi method in the form of operations on lists using the BSF-M representation. Let us define $L_{Map}$ as a list of

numbers of the matrix $C$ rows ordered in an arbitrary way:

$$L_{Map} = [j_1, \ldots, j_n], \quad (12)$$

where $j_k \in \{1, \ldots, n\}$ for $k = 1, \ldots, n$. For an arbitrary $x \in \mathbb{R}^n$, Let us define function $F_x : \{1, \ldots, n\} \to \mathbb{R}^n$:

$$F_x(i) = d_i + \sum_{j=1}^{n} c_{ij} x_j \quad (13)$$

for all $i \in \{1, \ldots, n\}$. In other words, the function $F_x(i)$ calculates the $i$-th coordinate of the next approximation.

The *Jacobi-M algorithm* consists of the following steps:

Step 1. $k := 0$; $x^{(0)} := d$; $L_{Map} := [1, \ldots, n]$.

Step 2. $x^{(k+1)} := Map(F_{x^{(k)}}, L_{Map})$.

Step 3. If $\|x^{(k+1)} - x^{(k)}\|^2 < \varepsilon$, go to Step 5.

Step 4. $k := k + 1$; go to Step 2.

Step 5. Stop.

The BSF model assumes that the algorithm is executed by a computer system consisting of one master-node and $K$ worker-nodes $(K > 0)$. For simplicity, we suppose from now on that

$$n = mK \quad (14)$$

for some $m \in \mathbb{N}$. During the initialization of the iterative process, the initial data of the problem (matrix $A$ and vector $b$) is copied to all nodes. Step 1 of the Jacobi-M algorithm is performed by both the master-node and the worker-nodes. Step 2 (*Map*) is performed only by the worker-nodes. The algorithm is executed in parallel according to Algorithm 1. Steps 3-4 are performed only on the master-node.

VII. ANALYTICAL STUDY OF THE JACOBI-M ALGORITHM

The BSF model assumes that all arithmetic operations (addition and multiplication) as well as comparison operation of floating-point numbers take the same time, which we denote as $\tau_{op}$. To perform the scalability analysis of the Jacobi-M algorithm, let us introduce the following notation:

$c_s$ – the quantity of real numbers transferred from the master to one worker;

$c_{Map}$ – the quantity of arithmetic operations performed in the *Map* step (Step 2 of the algorithm);

$c_r$ – the quantity of real numbers transferred from one worker to the master;

$c_p$ – the quantity of arithmetic operations performed by the master in Step 3 of the algorithm.

Let us calculate these values. At the beginning of the iteration, the master sends to each worker the current approximation $x^{(k)}$, which is a vector of length $n$. Consequently,

$$c_s = n. \quad (15)$$

Let us calculate the number of arithmetic operations performed in the *Map* step. For each element of the list $L_{Map}$, one vector is calculated by equation (13). Thus, it takes $2n$ floating-point operations. Multiplying this number by the number of rows in the matrix $C$, we get

$$c_{Map} = 2n^2. \quad (16)$$

In accordance with (14), each worker-node calculates $m$ coordinates of the next approximation, which must be transferred to the master-node. Hence,

$$c_r = m. \quad (17)$$

Execution of Step 3 requires $(2n + 2)$ operations. Hence, we obtain the following equation:

$$c_p = 2n + 2. \quad (18)$$

Let us assume $\tau_{op}$ — the time it takes the worker to perform one arithmetic operation or comparison operation, $\tau_{tr}$ — the time it takes to transfer a single float number across the network excluding latency. Then, we get the following values for the cost metrics of the BSF model in the case of the Jacobi-M algorithm:

$$t_s = \tau_{tr} n; \quad (19)$$

$$t_w = 2\tau_{op} n^2; \quad (20)$$

$$t_R = \tau_{tr} n; \quad (21)$$

$$t_p = 2\tau_{op}(n+1). \quad (22)$$

Equation (19), obtained on the basis of (15), gives an estimation of the time $t_s$ spent by the master to transfer a message to one worker excluding latency. Equation (20) obtained using equation (16) allows to estimate $t_w$, which denotes the total time spent by the workers for computations on local data. To obtain the estimation of the time $t_R$ spent by the master to transfer a message to all of worker-nodes excluding latency, $c_r$ must be multiplied by the number of workers: $t_R = K \cdot c_r$. Equation (22) obtained using equation (18) calculates the time $t_p$ spent by the master on the checking of stopping criterion. On the basis of equations (5) – (7) and (19) – (22), we obtain the following equations for speedup, parallel efficiency and the upper bound of scalability of the Jacobi-M algorithm:

$$a_{Jacobi-M}(K) = \\ = \frac{K(2L + \tau_{tr}n + \tau_{tr}n + \tau_{op} \cdot (2n+2) + \tau_{op} 2n^2)}{K^2(2L + \tau_{tr}n) + K(\tau_{tr}n + \tau_{op} \cdot (2n+2)) + \tau_{op} 2n^2}, \quad (23)$$

$$e_{Jacobi-M}(K) = \\ = \frac{2L + \tau_{tr}n + \tau_{tr}n + \tau_{op} \cdot (2n+2) + \tau_{op} 2n^2}{K^2(2L + \tau_{tr}n) + K(\tau_{tr}n + \tau_{op} \cdot (2n+2)) + \tau_{op} 2n^2}, \quad (24)$$

$$P_{Jacobi-M} = \sqrt{\frac{\tau_{op} \cdot 2n^2}{2L + \tau_{tr}n}}. \quad (25)$$

Let us simplify equation (25). When $n \to \infty$ we have

$$2L + \tau_{tr}n = O(n). \quad (26)$$

Substituting the right-hand side of equation (26) in (25), we obtain

$$P_{Jacobi-M} = \sqrt{\frac{O(n^2)}{O(n)}},$$

which is equivalent to

$$P_{Jacobi-M} = \sqrt{O(n)}. \quad (27)$$

Thus, the upper bound of the scalability $P_{Jacobi-M}$ of the Jacobi-M algorithm grows in proportion to the square root of the dimension of the problem.

VIII. JACOBI-MR ALGORITHM

The Jacobi-MR algorithm implements the Jacobi method in the form of operations on lists with the use of high-order functions *Map* and *Reduce*. Let us de define $L_{Map}$ as a list of the column numbers of the matrix $C$ ordered in an arbitrary way:

$$L_{Map} = [i_1, \ldots, i_n], \quad (28)$$

where $i_k \in \{1, \ldots, n\}$ for $k = 1, \ldots, n$. For an arbitrary $x \in \mathbb{R}^n$, let us define the function $F_x : \{1, \ldots, n\} \to \mathbb{R}^n$:

$$F_x(j) = (x_j c_{1j}, \ldots, x_j c_{nj}) \quad (29)$$

for all $j \in \{1, \ldots, n\}$. Informally, the function multiplies the $j$-th column of the matrix $C$ by the $j$-th coordinate of the vector $x$. For an arbitrary $x \in \mathbb{R}^n$, we define the list $L_{Reduce}^{(x)} \subset \mathbb{R}^n$ as follows:

$$L_{Reduce}^{(x)} = [F_x(j_1), \ldots, F_x(j_n)]. \quad (30)$$

The list $L_{Reduce}^{(x)}$ includes the matrix $C$ columns multiplied by the corresponding coordinate of vector $x$ taken in the order determined by the list $L_{Map}$. Thus, the list $L_{Reduce}^{(x)}$ can be obtained by applying to the list $L_{Map}$ the high-order function *Map* using as a parameter the function $F_x$:

$$L_{Reduce}^{(x)} = Map(F_x, L_{Map}). \quad (31)$$

Let us define the binary associative operation $\oplus : \mathbb{R}^n \times \mathbb{R}^n \to \mathbb{R}^n$ in the following way:

$$x \oplus y = x + y \quad (32)$$

for any $x, y \in \mathbb{R}^n$. In this case, the operation $\oplus$ performs the composition of vectors. Then, for the current approximation $x^{(k)}$, the next approximation $x^{(k+1)}$ can be obtained by applying to the list $L_{Reduce}^{(x^{(k)})}$ the high-order function *Reduce* using the operation $\oplus$ as a parameter and by adding the vector $d$ to the result of the *Reduce* operation.

$$x^{(k+1)} = Reduce\left(\oplus, L_{Reduce}^{(x^{(k)})}\right) + d. \quad (33)$$

*The Jacobi-MR algorithm* consists of the following steps:

Step 1. $k := 0$; $x^{(0)} := d$; $L_{Map} := [1, \ldots, n]$.

Step 2. $L_{Reduce}^{(x^{(k)})} := Map(F_{x^{(k)}}, L_{Map})$.

Step 3. $x^{(k+1)} := Reduce\left(\oplus, L_{Reduce}^{(x^{(k)})}\right)$.

Step 4. $x^{(k+1)} := x^{(k+1)} + d$.

Step 5. If $\|x_{k+1} - x_k\|^2 < \varepsilon$, go to Step 7.

Step 6. $k := k+1$; go to Step 2.

Step 7. Stop.

During the initialization of the iterative process, the initial data of the problem (matrix $A$ and vector $b$) is copied to all nodes. Step 1 of the Jacobi-MR algorithm is performed by both the master-node and the worker-nodes. Step 2 (*Map*) is performed only by the worker-nodes. The algorithm is executed in parallel according to Algorithm 2. Step 3 (*Reduce*) is performed on the worker-nodes and partially on the master-node. Steps 4-6 are performed only on the master-node.

## IX. ANALYTICAL STUDY OF THE JACOBI-MR ALGORITHM

To perform the scalability analysis of the Jacobi-M algorithm during one iteration, let us introduce the following notation:

$c_s$ – the quantity of real numbers transferred from the master to one worker;

$c_{Map}$ – the quantity of arithmetic operations performed in the *Map* step (Step 2 of the algorithm);

$c_a$ – the number of arithmetic operations necessary to calculate a composition of two vectors;

$c_r$ – the quantity of real numbers transferred from one worker to the master;

$c_p$ – the quantity of arithmetic operations performed by the master in Steps 4 and 5 of the algorithm.

Let us calculate these values. At the beginning of the iteration, the master sends to each worker the current approximation $x^{(k)}$, which is a vector of length $n$. Consequently,

$$c_s = n. \quad (34)$$

In the *Map* step, the columns of the matrix $C$ ordered by the numbers from the list $L_{Map}$ are multiplied by the corresponding coordinates of the vector $x^{(k)}$. This requires $n^2$ arithmetic operations. Consequently,

$$c_{Map} = n^2. \quad (35)$$

Composition of two vectors of length $n$ requires $n$ arithmetic operations. Consequently,

$$c_a = n. \quad (36)$$

The vector calculated by each worker in *Reduce* step must be transferred to the master. Hence,

$$c_r = n. \quad (37)$$

The execution of Step 4 requires $n$ arithmetic operations. The execution of Step 5 requires $3n-1$ arithmetic operations and one comparison operation. Hence, we obtain the following equation:

$$c_p = 3n. \quad (38)$$

We still assume that $\tau_{op}$ denotes the time it takes for a worker to perform one arithmetic operation, and $\tau_{tr}$ – the time it takes to transfer a single float number across the network excluding latency. Therefore, we obtain the following values for the cost parameters of the Jacobi-MR algorithm:

$$t_s = \tau_{tr} n; \quad (39)$$

$$t_w = c_{Map} + (m-1)c_a = \tau_{op}\left(n^2 + n(n-K)\right); \quad (40)$$

$$t_r = \tau_{tr} n; \quad (41)$$

$$t_a = \tau_{op} n; \quad (42)$$

$$t_p = 3\tau_{op} n. \quad (43)$$

Equation (39) obtained on the basis of (34) gives an estimation of the time $t_s$ spent by the master to transfer a message to one worker excluding latency. Equation (40) is obtained using equations (14), (35) and (36). Indeed, in accordance with the BSF cost metrics, $t_w$ denotes the summarized time costs which are spent by the workers to process local data. In accordance with the technique described in Section 2 (see Algorithm 2), the parallelization of Step 3 (*Reduce*) of the Jacobi-MR algorithm assumes every worker to process a sublist of length $m$. In that way, each worker has to produce $(m-1)$ compositions of vectors of length $n$. It requires $(m-1)c_a$ arithmetic operations. Hence, on the basis of equation (1), the total number of operations performed by the workers in the *Reduce* step is equal to $n(n-K)$. Adding to this the right-hand side of equation (35) and multiplying the result by $\tau_{op}$, we obtain the final equation (40) for $t_w$. Equation (41) obtained on the basis of (37) gives an estimation of the time $t_r$ spent by the master to receive a message from one worker excluding latency. Equation (42) obtained on the basis of (36) calculates the time $t_a$ spent by the master to perform its part of the *Reduce* step. Finally, equation (43) obtained on the basis of (38) calculates the time spent by the master for computing the next approximation and checking the stopping criterion.

Substituting the values of the right-hand sides of equations (39) – (43) into equation (8), we obtain the following equation for estimating the speedup of the Jacobi-MR algorithm:

$$a_{Jacobi-MR}(K) = \frac{2(L+\tau_{tr}n) + \tau_{op}n(3n-K+3)}{K\left(2(L+\tau_{tr}n)+\tau_{op}n\right)+3\tau_{op}n(n/K+1)}. \quad (44)$$

Equation (9) gives an estimation of the parallel efficiency of the Jacobi-MR algorithm:

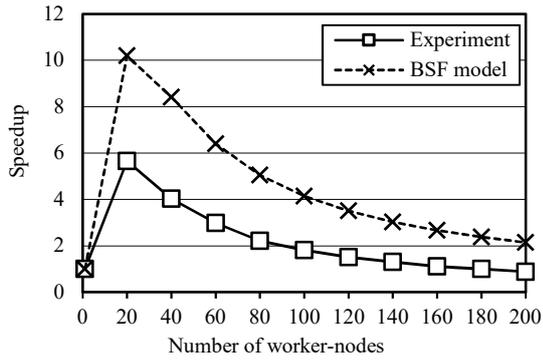

Fig. 1. Speedup of Jacobi-M algorithm for $n = 1500$.

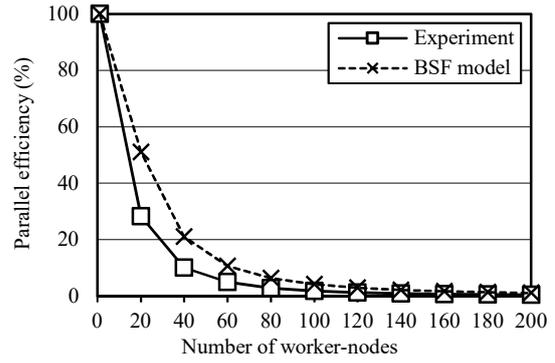

Fig. 5. Parallel Efficiency of Jacobi-M algorithm for $n = 1500$.

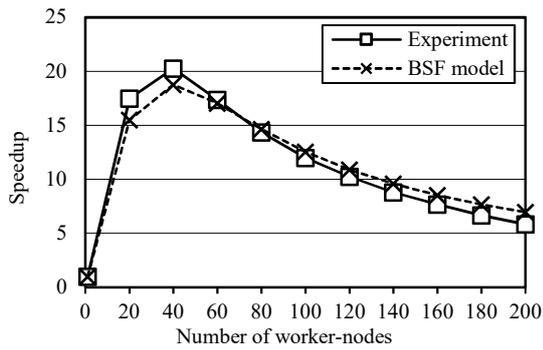

Fig. 2. Speedup of Jacobi-M algorithm for $n = 5000$.

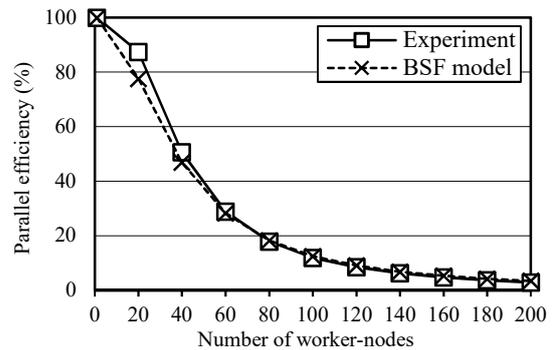

Fig. 6. Parallel Efficiency of Jacobi-M algorithm for $n = 5000$.

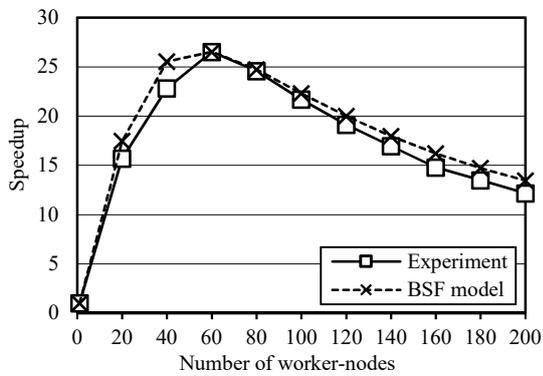

Fig. 3. Speedup of Jacobi-M algorithm for $n = 10000$.

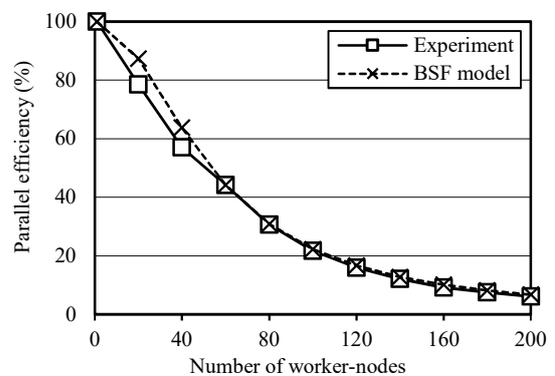

Fig. 7. Parallel Efficiency of Jacobi-M algorithm for $n = 10000$.

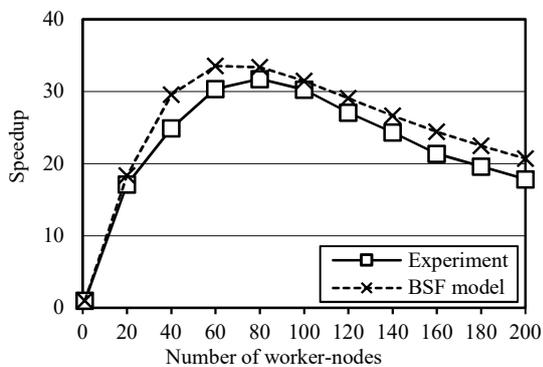

Fig. 4. Speedup of Jacobi-M algorithm for $n = 16000$.

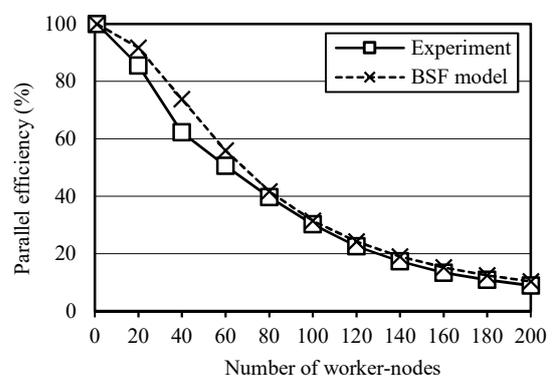

Fig. 8. Parallel Efficiency of Jacobi-M algorithm for $n = 16000$.

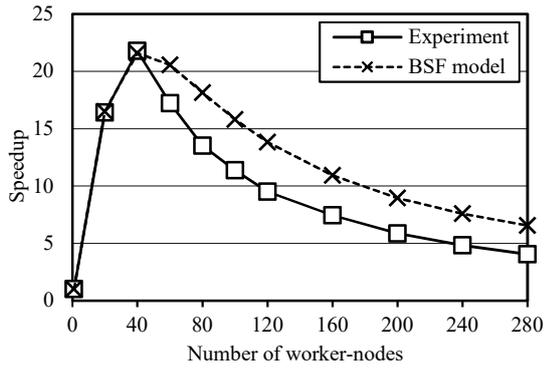

Fig. 9. Speedup of Jacobi-MR algorithm for $n = 1500$.

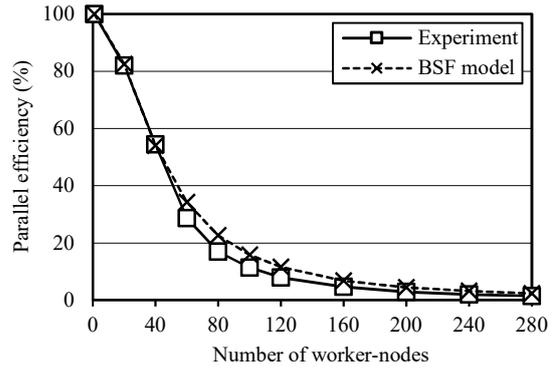

Fig. 13. Parallel Efficiency of Jacobi-MR algorithm for $n = 1500$.

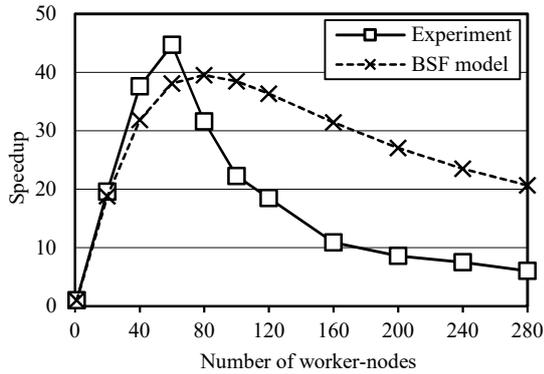

Fig. 10. Speedup of Jacobi-MR algorithm for $n = 5000$.

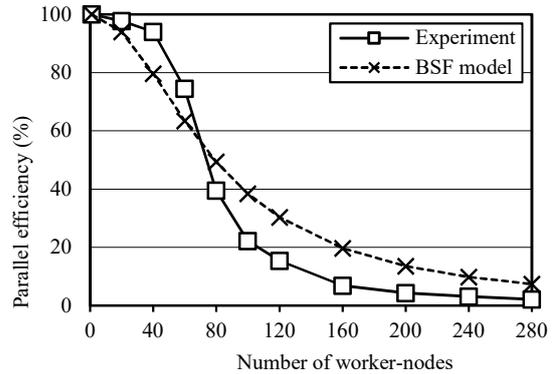

Fig. 14. Parallel Efficiency of Jacobi-MR algorithm for $n = 5000$.

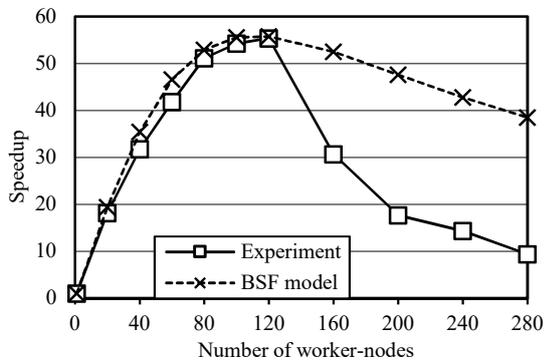

Fig. 11. Speedup of Jacobi-MR algorithm for $n = 10000$.

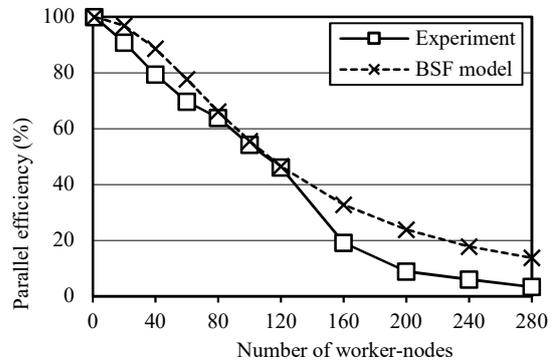

Fig. 15. Parallel Efficiency of Jacobi-MR algorithm for $n = 10000$.

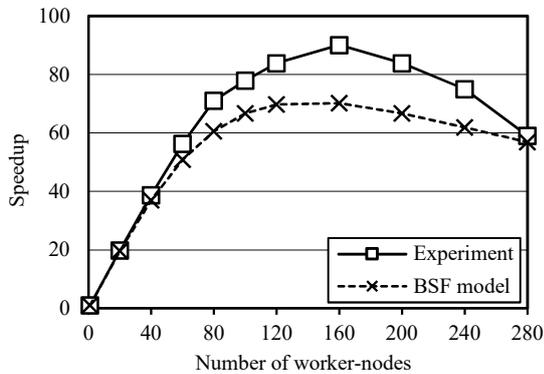

Fig. 12. Speedup of Jacobi-MR algorithm for $n = 16000$.

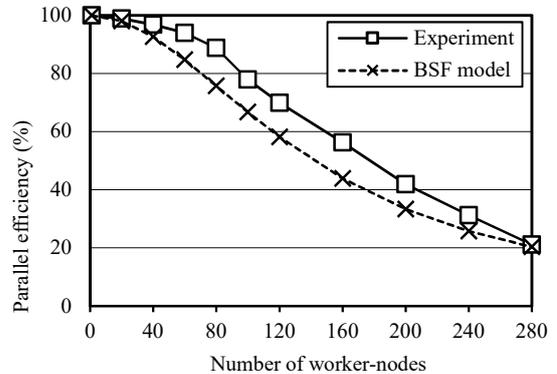

Fig. 16. Parallel Efficiency of Jacobi-MR algorithm for $n = 16000$.

$$e_{Jacobi-MR}(K) = \frac{2(L+\tau_{tr}n)+\tau_{op}n(3n-K+3)}{K^2\left(2(L+\tau_{tr}n)+\tau_{op}n\right)+3\tau_{op}n(n+K)}. \quad (45)$$

The upper bound of the Jacobi-MR algorithm scalability is obtained by substituting the values of the right-hand sides of equations (39) – (43) into equation (10):

$$P_{Jacobi-MR} = \sqrt{\frac{\tau_{op}n(3n-K)}{2(L+\tau_{tr}n)+\tau_{op}n}}. \quad (46)$$

Let us simplify equation (46). For $n \to \infty$, we have

$$2(L+\tau_{tr}n)+\tau_{op}n = O(n). \quad (47)$$

From (14), it follows that $K \le n$. Then, for $n \to \infty$:

$$\tau_{op}n(3n-K) = O(n^2). \quad (48)$$

Substituting the right-hand sides of equations (47) and (48) into (46), we obtain

$$P_{Jacobi-MR} = \sqrt{\frac{O(n^2)}{O(n)}},$$

which is equivalent to

$$P_{Jacobi-MR} = \sqrt{O(n)}. \quad (49)$$

Thus, the upper bound of the scalability of the Jacobi-MR algorithm on lists also grows in proportion to the square root of the dimension of the problem $n$.

## X. COMPUTATIONAL EXPERIMENTS

To verify the results obtained analytically, we implemented of the Jacobi-M and Jacobi-MR algorithms in C++ using the BSF program skeleton and the MPI parallel programming library. The source codes of these programs are freely available on Github, at https://github.com/nadezhda-ezhova/Jacobi-M and https://github.com/nadezhda-ezhova/Jacobi-MR respectively. To carry out the experiments, we used a scalable system of linear equations (11) having the following coefficient matrix $A$ and the vector of constant terms $b$:

$$A = \begin{pmatrix} 1 & 1 & \cdots & 1 \\ 1 & 2 & \ddots & \vdots \\ \vdots & \ddots & \ddots & 1 \\ 1 & \cdots & 1 & n \end{pmatrix}, \quad b = \begin{pmatrix} n \\ n+1 \\ \vdots \\ 2n-1 \end{pmatrix}.$$

We investigated the speedup and parallel efficiency of the Jacobi-M and Jacobi-MR algorithms on the supercomputer "Tornado SUSU" [21]. The calculations were performed for the dimensions 1 500, 5 000, 10 000 and 16 000. At the same time, we plotted the curves of speedup and parallel efficiency for these dimensions using equations (23), (24), (44) and (45). For this, the following values in seconds were determined experimentally: $L = 1.5 \cdot 10^{-5}$, $\tau_{op} = 2.9 \cdot 10^{-8}$ and $\tau_{tr} = 1.9 \cdot 10^{-7}$. The results are presented in Fig. 1–16. In all cases, the analytical estimations are very close to experimental ones. Moreover, the performed experiments show that the upper bound of the BSF-program scalability increases proportionally to the square root of problem dimension. It was analytically predicted using the equations (27) and (49).

## XI. CONCLUSION

In this paper, the scalability and parallel efficiency of the iterative Jacobi-M and Jacobi-MR algorithm used to solve large-scale linear equation systems on multiprocessors with distributed memory were investigated. To do this, we used the BSF (Bulk Synchronous Farm) parallel computation model based on the "master-worker" paradigm. The BSF-implementations of the Jacobi-M and Jacobi-MR algorithms in the form of operations on lists using high-order functions *Map* and *Reduce* defined in the Bird–Meertens formalism were described. The scalability upper bounds of the BSF-implementations of Jacobi-M and Jacobi-MR algorithms were obtained. Also, the equations for estimating the speedup and parallel efficiency of the Jacobi-M and Jacobi-MR algorithm on lists was obtained. The implementations of the Jacobi-M and Jacobi-MR algorithm in C++ language using the BSF algorithmic skeleton and the MPI parallel programming library were performed. These implementations are freely available on Github, at https://github.com/nadezhda-ezhova/Jacobi-M and https://github.com/nadezhda-ezhova/Jacobi-MR respectively. The large-scale experiments were conducted on a cluster computing system to obtain the actual speedup and parallel efficiency curves for the linear equation systems having the following dimensions: 1 500, 5 000, 10 000, 16 000. The results of the experiments showed that the BSF model predicts the upper bound of the scalability of the Jacobi-M and Jacobi-MR algorithms on lists with high accuracy. As future research directions, we intend to design and implement an integrated development environment that provides comprehensive facilities to computer programmers for developing BSF-programs.

ACKNOWLEDGMENT

The study was partially supported by the RFBR according to research project No. 17-07-00352-a, by the Government of the Russian Federation according to Act 211 (contract No. 02.A03.21.0011) and by the Ministry of Science and Higher Education of the Russian Federation (government order 2.7905.2017/8.9).

REFERENCES

[1] K. Borodulin, G. Radchenko, A. Shestakov, L. Sokolinsky, A. Tchernykh, and R. Prodan, "Towards Digital Twins Cloud Platform," in Proceedings of the 10th International Conference on Utility and Cloud Computing - UCC '17, 2017, pp. 209–210.

[2] G. Bilardi and A. Pietracaprina, "Models of Computation, Theoretical," in Encyclopedia of Parallel Computing, Boston, MA: Springer US, 2011, pp. 1150–1158.

[3] J. F. JaJa, "PRAM (Parallel Random Access Machines)," in Encyclopedia of Parallel Computing, Boston, MA: Springer US, 2011, pp. 1608–1615.

[4] L. G. Valiant, "A bridging model for parallel computation," Commun. ACM, vol. 33, no. 8, pp. 103–111, 1990.

[5] D. Culler et al., "LogP: towards a realistic model of parallel computation," in Proceedings of the fourth ACM SIGPLAN symposium on Principles and practice of parallel programming - PPOPP'93, 1993, pp. 1–12.

[6] M. Forsell and V. Leppanen, "An extended PRAM-NUMA model of computation for TCF programming," in Proceedings of the 2012 IEEE 26th International Parallel and Distributed Processing Symposium Workshops, IPDPSW 2012, 2012, pp. 786–793.

[7] A. V. Gerbessiotis, "Extending the BSP model for multi-core and out-of-core computing: MBSP," Parallel Comput., vol. 41, pp. 90–102, 2015.

[8] F. Lu, J. Song, and Y. Pang, "HLognGP: A parallel computation model for GPU clusters," Concurr. Comput. Pract. Exp., vol. 27, no. 17, pp. 4880–4896, Dec. 2015.

[9] L. B. Sokolinsky, "Analytical Estimation of the Scalability of Iterative Numerical Algorithms on Distributed Memory Multiprocessors," Lobachevskii J. Math., vol. 39, no. 4, pp. 571–575, 2018.

[10] L. M. Silva and R. Buyya, "Parallel programming models and paradigms," in High Performance Cluster Computing: Architectures and Systems. Vol. 2., 1999, pp. 4–27.

[11] F. Darema, "SPMD Computational Model," in Encyclopedia of Parallel Computing, Boston, MA: Springer US, 2011, pp. 1933–1943.

[12] S. Sahni and G. Vairaktarakis, "The master-slave paradigm in parallel computer and industrial settings," J. Glob. Optim., vol. 9, no. 3–4, pp. 357–377, Dec. 1996.

[13] I. Sokolinskaya and L. B. Sokolinsky, "Scalability Evaluation of NSLP Algorithm for Solving Non-Stationary Linear Programming Problems on Cluster Computing Systems," Supercomput. RuSCDays 2017. Commun. Comput. Inf. Sci., vol. 793, pp. 40–53, 2017.

[14] I. M. Sokolinskaya and L. B. Sokolinsky, "Scalability Evaluation of Cimmino Algorithm for Solving Linear Inequality Systems on Multiprocessors with Distributed Memory," Supercomput. Front. Innov., vol. 5, no. 2, pp. 11–22, 2018.

[15] M. I. Cole, "Parallel programming with list homomorphisms," Parallel Process. Lett., vol. 05, no. 02, pp. 191–203, Jun. 1995.

[16] H. Rutishauser, "The Jacobi Method for Real Symmetric Matrices," in Handbook for Automatic Computation, vol 2. Linear Algebra, Bauer F.L., Ed. Berlin, Heidelberg: Springer Berlin Heidelberg, 1971, pp. 202–211.

[17] C. G. J. Jacobi, "Ueber eine neue Auflösungsart der bei der Methode der kleinsten Quadrate vorkommenden lineären Gleichungen," Astron. Nachrichten, vol. 22, no. 20, pp. 297–306, 1845.

[18] H. H. Goldstine, F. J. Murray, and J. von Neumann, "The Jacobi Method for Real Symmetric Matrices," J. ACM, vol. 6, no. 1, pp. 59–96, Jan. 1959.

[19] X. I. A. Yang and R. Mittal, "Acceleration of the Jacobi iterative method by factors exceeding 100 using scheduled relaxation," J. Comput. Phys., vol. 274, pp. 695–708, Oct. 2014.

[20] J. E. Adsuara, I. Cordero-Carrión, P. Cerdá-Durán, and M. A. Aloy, "Scheduled Relaxation Jacobi method: Improvements and applications," J. Comput. Phys., vol. 321, pp. 369–413, Sep. 2016.

[21] P. S. Kostenetskiy and A. Y. Safonov, "SUSU Supercomputer Resources," in Proceedings of the 10th Annual International Scientific Conference on Parallel Computing Technologies (PCT 2016). CEUR Workshop Proceedings. Vol. 1576, 2016, pp. 561–573.